\documentstyle[twocolumn,prb,aps,epsf,rotate]{revtex}
\large
\begin{document}
\epsfverbosetrue
\draft
\twocolumn[\hsize\textwidth\columnwidth\hsize\csname @twocolumnfalse\endcsname
\title
{X-ray absorption and resonant inelastic x-ray scattering in the rare earths}
\author{Michel van Veenendaal and Robert Benoist}
\address{European Synchrotron Radiation Facility,
B.P. 220, F-38043 Grenoble C\'{e}dex, France.}
\date{\today} 
\maketitle 

\begin{abstract}
This paper makes a comparison between x-ray absorption (XAS) and 
resonant inelastic x-ray scattering (RIXS) in the
rare earths. Atomic calculations
are given for $2p\rightarrow 4f$ and $2p\rightarrow 5d$ XAS. 
The latter calculation includes the contraction and expansion of the
$5d$ orbitals resulting from the complete
exchange  interaction with the $4f$ electrons. 
The radiative decay of the XAS final states is described for
the situations where the core hole created in the absorption process
is filled by a valence electron or by an electron from a shallower
core level. RIXS spectra, 
$4f^n\rightarrow \underline{3d}4f^{n+1}\rightarrow 4f^n$, integrated 
 over the outgoing photon energy (fluorescence yield) are compared with
$3d\rightarrow 4f$ XAS. Sum rules related to XAS and RIXS and their
applicability are discussed. 
\end{abstract}
]
\section{INTRODUCTION}
Experiments on rare earth systems have made 
an important contribution to the development
of magnetic x-ray dichroism and resonant magnetic scattering. 
Both circular\cite{GSGd} and linear\cite{GvdL} magnetic x-ray dichroism
have been observed for the first time in rare earths.
The first satisfactory explanation of resonant magnetic scattering 
was done for Holmium metal.\cite{Gibbs,Hannon}
The electronic structure of rare earths is determined by the 
interaction between electrons in the localized $4f$ orbitals and in 
the broad $5d$ band. Spectroscopies involving the $4f$ shell
can usually be successfully described by atomic multiplet theory.
But whereas, e.g., magnetic x-ray dichroism at the $M_{45}$ edges
is rather well understood, the $L_{23}$ spectra, dominated by the dipolar
$2p\rightarrow 5d$ transitions,
pose more problems. In the interpretation one has to take
into account two effects. 

First, one observes pre-edge features\cite{PCAl,PCGd}
that are weak in the isotropic spectra but have a strong
circular dichroic signal.
That these structures arise from
quadrupolar transitions into the $4f$ shell has been established
by resonant elastic $\sigma\mathopen{\rightarrow}\pi$ 
x-ray scattering,\cite{Gibbs,Hannon}
by the observation of a non-dipolar
angular dependence of the circular dichroic XAS,\cite{Harm,AngDep}
by resonant Raman spectroscopy,\cite{Ham,Kri} and by partial deconvolution
of the life time broadening.\cite{Loef} 

Second, the finite integrated intensity of the circular dichroism
of the $2p\rightarrow 5d$ transitions results, not only
from the polarization of the $5d$ electrons in the ground state, but
also from a dependence of the $2p$-$5d$ radial matrix elements
on the direction of the $5d$ moment 
relative to that of the $4f$.\cite{Harm,Har}
Therefore, band effects\cite{Harm} and the 
full $df$-Coulomb interaction\cite{JoIm,MvVl23} have to be included  
in the interpretation of the spectral line shape and
the variations in the $L_{23}$ circular dichroic branching ratios.

A complication in the interpretation of RIXS is that the deexcitation
cannot be simply decoupled from the absorption step as can be done
when the excitation is far above threshold. Furthermore, the
decay is different for transitions 
between two core levels or between the valence shell and a core level.

For systems where a detailed description of the spectral line shape
is complex, useful results can still be obtained from statistical
methods. For XAS and x-ray scattering sum rules exist
that relate the integrated intensities to ground state 
properties.\cite{Thsum,PCsum,Luo,PCRRS}
These sum rules rely on a number of approximations.
In general  a constant radial matrix element
is required. For sum rules that consider the spin-orbit
manifolds separately one has to assume that the edges can be 
distinguished by the $j$-value of the core hole. For fluorescence yield
no exact sum rules have been derived so far and the application 
of XAS sum rules requires numerical validation.\cite{MvVfl}
In this paper we address the applicability of the sum rules.

The paper is divided as follows. First, we show the 
separation of the geometric and electronic part for XAS and 
RIXS. Section III gives a derivation
for the angular distributions for some common experimental
situations. Section IV is devoted to the 
spectral functions and their sum rules.
We describe the spectral line shape of the XAS spectra at the
$L_{23}$ edge. For the dipolar part we include the effects of
contraction and expansion of the $5d$ orbital by the $df$-Coulomb 
interaction. RIXS results are discussed for spectroscopies involving
the $4f$ shell. A comparison is made between the radiative decay in
spectroscopies where the intermediate state core hole is filled
by a valence electron or by an electron from a shallower core level.
We end with the conclusions.
\section{Intensities}
The intershell  transitions as a result of the absorption or emission
of a x-ray photon are described by 
\makebox{$H_{{\rm int}}= \frac{e}{2m}\{  {\rm{\bf p}} \cdot {\rm{\bf A}}
+ {\rm{\bf A}} \cdot {\rm{\bf p}} \}$}.
Expanding the vector potential in plane waves gives
\begin{eqnarray}
{\rm{\bf A}}= \sum_{{\rm{\bf k}}{\hbox{\boldmath{$\epsilon$}}}}
  \sqrt{ \frac{\hbar}{2 \omega \epsilon_0 \Omega}}
\left \{ \makebox{\boldmath {$ \epsilon$}} 
a_{{\rm{\bf k}}{\hbox{\boldmath{$\epsilon$}}}}
 {\rm e}^{i{\rm{\bf k}}{\rm{\bf r}} } + {\rm h.c.} \right \} ,
\end{eqnarray}
where $\Omega$ is  a normalization volume and
 $a_{{\rm{\bf k}}{\makebox{\boldmath{$\epsilon$}}}}$
annihilates a photon with momentum ${\rm{\bf k}}$ and polarization
vector $\boldmath{\epsilon}$. The absorption intensity is then
given by Fermi's golden rule:
\begin{eqnarray}
I({\hat {\rm{\bf k}}}\makebox{\boldmath {$ \epsilon$}}\omega)=
 \frac{2 \pi}{\hbar} \sum_n |\langle n | H_{{\rm int}} |g \rangle |^2
 \delta(\omega + E_g -E_n) \rho_{\omega},
\end{eqnarray}
with the density of oscillators given by 
$\rho_{\omega}=  \frac{\Omega}{8 \pi^3} 
\frac{\omega^2}{\hbar c^3}$. The different multipoles of the spectrum are
obtained by expanding the plane wave 
in Bessel functions and spherical harmonics, i.e,
\begin{eqnarray}
 {\rm e}^{i{\rm{\bf k}}{\rm{\bf r}} } =
\sum_{t} [t] i^t j_t(kr) 
{\hat {\rm{\bf k}}}^{(t)} \cdot 
{\hat {\rm{\bf r}}}^{(t)},
\end{eqnarray}
with $[a\dots b]=(2a+1)\dots (2b+1)$.
We use here the shorthands 
${\rm{\bf k}}^{(l)}=k^l {\rm{\bf C}}^l({\hat {\rm{\bf k}}})$
and for spherical tensors of rank one:
${\rm{\bf k}}={\rm{\bf k}}^{(1)}$. Note that ${\hat k}=1$.
For $kr\mathopen{\ll} 1$ one has $j_t(kr)\mathopen{\cong}(kr)^t/[t]!!$.
With the use of the orthogonality relation one then obtains
\begin{eqnarray}
{\rm{\bf p}} \cdot {\makebox{\boldmath {$ \epsilon$}}}
{\rm e}^{i{\rm{\bf k}}{\rm{\bf r}} } =
\sum_{tQ} [tQ]   \frac{i^t}{[t]!!}
[{\rm{\bf p}},
{\rm{\bf r}}^{(t)} ]^Q \mathopen{\cdot }
[\makebox{\boldmath {$ \epsilon$}} ,
{\rm{\bf k}}^{(t)} ]^Q ,
\end{eqnarray}
where the tensor couplings are defined in the Appendix.
It is convenient to define the operators
\begin{eqnarray}
{\rm {\bf V}}^{tQ}=
\frac{b_{tQ}(k)}{2i m\omega } \{ ~
[{\rm{\bf p}},{\rm{\bf r}}^{(t)} ]^Q + (-1)^Q
[ {\rm{\bf r}}^{(t)},{\rm{\bf p}} ]^Q
~ \} ,
\label{opers}
\end{eqnarray}
where the factors $b_{tQ}(k)$ will be chosen in such a way that 
nicely defined operators are obtained, see Table \ref{operators}.
For $tQ=01$ (electric dipole) and $b_{01}(k)=\sqrt{3}$ we have
\begin{eqnarray}
\langle n| {\rm {\bf V}}^{01} | g \rangle
&=& \frac{b_{01}(k)}{\sqrt{3}i m \omega }\langle n|{\rm{\bf p}} |g\rangle
= \frac{1}{\hbar \omega}
\langle n|[\frac{ p^2}{2 m},{\rm{\bf r}}] |g\rangle
\cong\langle n|{\rm{\bf r}}|g \rangle .
\nonumber
\end{eqnarray}
By using the definition for the outer product and  
$b_{11}(k)=\sqrt{\frac{3}{2}}k$,  we find that 
${\rm {\bf V}}^{11}$ is equal to $\frac{b_{11}(k)}{i m\omega }
[{\rm{\bf p}},{\rm{\bf r}}]^1
=\frac{\alpha}{2}\frac{a_0}{\hbar} {\rm {\bf L}}$,
with $\alpha$ the fine-structure
constant and $a_0$ the Bohr radius. 
${\rm {\bf V}}^{11}$ forms together with the $g_S {\rm {\bf S}}$ term
the magnetic dipole operator. Magnetic dipole transitions are 
about $(\alpha/2)^2$, i.e., five orders of magnitude smaller than
electric dipole transitions of the same wavelength.
Furthermore, we have for the electric quadrupole operator,
${\rm {\bf V}}^{12}\mathopen{=} {\rm{\bf r}}^{(2)}$,
that $b_{12}(k)=\sqrt{30}$.

The multipole expansion enables us to separate the 
absorption intensity into a geometric and an electronic 
\begin{table}
\caption{Relevant constants for electric dipole ($tQ$=01),
magnetic dipole (11), and electric quadrupole (12) transitions.
The factors $b_{tQ}(k)$ appear in the definition of the operators 
${\rm {\bf V}}^{tQ}$.
The $B_Q^2$ give the relative probabilities of dipolar and quadrupolar
transitions. The factors $D_{tQ}$ are used in the definition
of the angular dependence ${\rm {\bf T}}^{tQz}$.}
\begin{center}
\begin{tabular}{ccccccc}
 & $t$ & $Q$ & ${\rm {\bf V}}^{tQ}$ & $b_{tQ}(k)$ & 
$B_Q^2$ & $D_{tQ}$ \\ [1.0mm] 
\hline \\
el. dip. & 0 & 1 & ${\rm {\bf r}}$ 
& $\sqrt{3}$ & $\frac{1}{3}$ & 3 \\
magn. dip. & 1 & 1 & $\frac{\alpha}{2}\frac{a_0}{\hbar} {\rm {\bf L}}$ 
& $\sqrt{\frac{3}{2}}k$ & $\frac{1}{3}$ & $3\sqrt{2}$ \\
el. quad. & 1 & 2 & $ {\rm {\bf r}}^{(2)}$ 
& $-\sqrt{30}$ & $\frac{1}{15}(\frac{k}{2})^2$ & $-5\sqrt{2}$ \\
\end{tabular}
\end{center} 
\label{operators}
\end{table}
part, i.e.,
\begin{eqnarray}
I({\hat {\rm{\bf k}}}\makebox{\boldmath {$ \epsilon$}} \omega)
 &=&\frac{2\pi}{\hbar} N_{\omega} \sum_{tQn} B_{Q}^2 D_{tQ}^2 ~
| \langle n |
[\makebox{\boldmath {$ \epsilon$}}, {\rm{\bf k}}^{(t)} ]^Q 
\cdot {\rm {\bf V}}^{tQ}
|g \rangle |^2
\nonumber \\  &&\qquad\qquad \qquad\qquad \qquad 
\times \delta(\omega + E_g -E_n)
\nonumber \\ &=& \frac{2\pi}{\hbar}
N_{\omega} \sum_{tQz} B_{Q}^2 
~ {\rm {\bf T}}^{tQz} 
({\hat {\rm{\bf k}}}\makebox{\boldmath {$ \epsilon$}})
\cdot 
{\rm {\bf I}}^{tQz}(\omega) ,
\end{eqnarray} 
where cross terms between different $tQ$ values have been omitted.
The following factors have been defined:
$N_{\omega}=\frac{e^2 \omega^3}{16 \pi^3 \epsilon_0 c^3}$;
$B_Q= (\frac{k}{2})^{Q-1}([Q]!!)^{\mathopen{-}\frac{1}{2}}$
gives the relative transition
probability, where for the dipolar and quadrupolar contributions
one has $B_2^2/B_1^2= \frac{1}{5}(\frac{k}{2})^2$; and
$D_{tQ}= [tQ]k^t/([t]!! b_{tQ}(k)B_Q )$.
The different multipole spectra are given by
\begin{eqnarray}
{\rm {\bf I}}^{tQz}(\omega)= \frac{\Gamma}{\pi}
\sum_n \frac{1}{|{\cal E}_n|^2} {\rm {\bf I}}^{tQz} (gnng)
\end{eqnarray}
with ${\cal E}_n=\omega+E_g-E_n +i\frac{\Gamma}{2}$ where $\Gamma$ is
the lifetime broadening of the XAS final states; the different
combinations of the matrix elements are defined as
\begin{eqnarray}
{\rm {\bf I}}^{tQz}(abcd)=
[\langle a|({\rm {\bf V}}^{\dagger})^{tQ} |b\rangle ,
\langle c|{\rm {\bf V}}^{tQ} |d\rangle
]^z
n_{Qz}^{-1} ~ ,
\end{eqnarray}
where the normalization constants $n_{Qz}$ are defined in the Appendix.
The angular distribution is given by
\begin{eqnarray} 
{\rm {\bf T}}^{tQz}
({\rm{\bf {\hat k}}}\makebox{\boldmath {$ \epsilon$}})&=&
[z]  D_{tQ}^2 
[ [ \makebox{\boldmath {$ \epsilon$}}^*,
{\hat {\rm{\bf k}}}^{(t)} ]^{Q} ,
[\makebox{\boldmath {$ \epsilon$}}, 
{\hat {\rm{\bf k}}}^{(t)} ]^Q ]^z
n_{Qz} ~ .
\label{angle}
\end{eqnarray} 

The RIXS intensity is proportional to
\begin{eqnarray} 
I({\hat {\rm{\bf k}}}\makebox{\boldmath {$ \epsilon$}}\omega,
{\hat {\rm{\bf k}}}'\makebox{\boldmath {$ \epsilon$}}'\omega')
&\mathopen{=}&\frac{2\pi}{\hbar}
\sum_f \left |
\sum_n \frac{\langle f| H_{{\rm int}} | n \rangle
\langle n| H_{{\rm int}} | g \rangle}
{\omega +E_g -E_n + i \frac{\Gamma}{2}}
\right |^2
\nonumber \\ &&  \qquad\times  
\delta (\omega +E_g -\omega' -E_f)\rho_{\omega'}\rho_{\omega} .
\end{eqnarray} 
In the remainder of the paper we consider the different
multipole transitions separately;
to reduce the number of indices we remove the $tQ$ of the transitions.
In a way similar to that for XAS we find for the 
RIXS cross section with  $Q$-polar
excitation followed by a $Q'$-polar deexcitation
\begin{eqnarray}
I({\hat {\rm{\bf k}}}\makebox{\boldmath {$ \epsilon$}} \omega,
{\hat {\rm{\bf k}}}'\makebox{\boldmath {$ \epsilon$}}' \omega')
 &=& \frac{2\pi}{\hbar} N_{\omega'}  B_{Q'}^2 N_{\omega}  B_{Q}^2 
\nonumber \\ &\times&
\sum_{zz'r}
 {\rm {\bf T}}^{zz'r} (
{\rm{\bf {\hat k}}}\makebox{\boldmath {$ \epsilon$}},
{\rm{\bf {\hat k}}}'\makebox{\boldmath {$ \epsilon$}}')
\cdot 
{\rm {\bf I}}^{zz'r}(\omega,\omega') .
\label{RIXS}
\end{eqnarray} 
The electronic part has now become a tensor product of absorption
and emission
\begin{eqnarray} 
{\rm {\bf I}}^{zz'r}(\omega,\omega') &=&
[r]   \frac{\gamma}{\pi}
\sum_{nn'f} \frac{1}{|{\cal E}_f|^2} \frac{1}{{\cal E}_{n'}^* {\cal E}_n} 
\nonumber \\ &\times&
[ {\rm {\bf I}}^{z'} (n'ffn),{\rm {\bf I}}^z (gn'ng)]^r n_{zz'r}~ ,
\end{eqnarray} 
where ${\cal E}_f=\omega+E_g -\omega' -E_f +i\frac{\gamma}{2}$ with $\gamma$
the lifetime broadening of the x-ray inelastic scattering final
states.  
The angular dependence is given by
\begin{eqnarray} 
{\rm {\bf T}}^{zz'r}
({\rm{\bf {\hat k}}}\makebox{\boldmath {$ \epsilon$}},
{\rm{\bf {\hat k}}}'\makebox{\boldmath {$ \epsilon$}}')=
[{\rm {\bf T}}^{z'}
({\rm{\bf {\hat k}}}'\makebox{\boldmath {$ \epsilon$}}'),
{\rm {\bf T}}^{z}
({\rm{\bf {\hat k}}}\makebox{\boldmath {$ \epsilon$}}) ]^r 
n_{zz'r}^{-1}  ~ .
\end{eqnarray} 

\section{Angular dependence}
The tensors for the angular dependence are chosen in such a
way that ${\rm {\bf T}}^0 \mathopen{=}1$ for $tQ\mathopen{=}01,11,12$.
For electric
dipole transitions ${\rm {\bf T}}^z$ is ${\hat {\rm {\bf k}}}$ independent,
and the $\zeta=0$ component is, with  respect 
to the ${\hat {\rm {\bf Z}}}$-axis of our system,
\begin{eqnarray} 
T^{z}_0(\makebox{\boldmath {$ \epsilon$}})=
3 [z](-1)^{1+z} n_{Qz} n_{11z} 
U^{11z}(\makebox{\boldmath {$ \epsilon$}}^*,
\makebox{\boldmath {$ \epsilon$}},{\hat {\rm {\bf Z}}}) ,
\label{Tzdip}
\end{eqnarray} 
where the bipolar spherical harmonics\cite{VMK} are given by
$U^{xyz}({\rm {\bf a}},{\rm {\bf b}},{\rm {\bf c}})
=(-)^{y+z} {\underline n}_{xyz}^{-1} 
[{\rm {\bf a}}^{(x)} {\rm {\bf b}}^{(y)} ]^z 
\mathopen{\cdot} {\rm {\bf c}}^{(z)}$. The bipolar spherical harmonics
with $xyz$ relevant for dipolar transitions are given in Table \ref{bipolar};
expressions for
higher values of $xyz$ are given by Thole and Van der Laan.\cite{ThIII}
For general multipole transitions it is 
more convenient to recouple the angular dependence 
\begin{eqnarray} 
{\rm {\bf T}}^{z}
({\rm{\bf {\hat k}}}\makebox{\boldmath {$ \epsilon$}})
&=&  (-1)^{1+t+z} D_{tQ}^2 \sum_{xy} [xyz]
\left \{
\begin{array}{ccc}
1 & x & 1 \\
t & y & t \\
Q & z & Q
\end{array}
\right \}
\nonumber \\ &\times& 
[ [ \makebox{\boldmath {$ \epsilon$}}^*, \makebox{\boldmath {$ \epsilon$}}]^x ,
[{\hat {\rm{\bf k}}}^{(t)}, 
{\hat {\rm{\bf k}}}^{(t)} ]^y ]^z
n_{Qz} ~ .
\end{eqnarray} 
It is straightforward to show that for left and right circularly
polarized light one has
\begin{eqnarray} 
\frac{1}{2} (
[ \makebox{\boldmath {$ \epsilon$}}^*_{+} ,
\makebox{\boldmath {$ \epsilon$}}_{+}]^x
+ (-1)^{x+m} 
[ \makebox{\boldmath {$ \epsilon$}}^*_{-},
 \makebox{\boldmath {$ \epsilon$}}_{-}]^x )
\mathopen{=}
\left \{
\begin{array}{cc}
a_x {\underline n}_{11x} C^x({\hat {\rm{\bf k}}}) & m\mathopen{=}0 \\ [1.0mm]
0 & m\mathopen{=}1
\nonumber
\end{array} 
\right . ,
\end{eqnarray} 
with $a_x =1,\frac{2}{3}i,-\frac{1}{2}$ for $x=0,1,2$. 
After that one finds with the use of the theorem
to couple two spherical harmonics of the same vector
(see Appendix) that the angular dependence is
a function of ${\rm {\bf C}}^z({\hat {\rm{\bf k}}})$,\cite{PCAl}
\begin{eqnarray} 
\frac{1}{2} (
{\rm {\bf T}}^z( {\hat {\rm{\bf k}}} \makebox{\boldmath {$ \epsilon$}}_{+} )
+ (-1)^{z+m}
{\rm {\bf T}}^z( {\hat {\rm{\bf k}}} \makebox{\boldmath {$ \epsilon$}}_{-} ))
\mathopen{=}
\left \{
\begin{array}{cc}
{\cal T}_z {\rm {\bf C}}^z({\hat {\rm{\bf k}}}) & m\mathopen{=}0 \\ [1.0mm]
0 & m\mathopen{=}1
\end{array}
\right .  
\label{Tzeqn}
\end{eqnarray} 
where the coefficients 
\begin{eqnarray} 
{\cal T}_z&\mathopen{=}&(-)^{1+t+z} \sum_{xy} [xyz] a_x D_{tQ}^2
\left \{
\begin{array}{ccc}
1 & x & 1 \\
t & y & t \\
Q & z & Q
\end{array}
\right \}
\nonumber 
n_{Qz} {\underline n}_{11x} n_{tty} n_{xyz}
\end{eqnarray} 
are given in Table \ref{Tz}.
\section{Spectra and sum rules}
\subsection{XAS into the $4f$ shell}  
Here we consider two absorption edges often studied: the $L_{23}$-edges,
corresponding to transitions from the $2p$ orbital into the valence shell
 and the $M_{45}$-edges that are 
\begin{table}
\caption{Bipolar spherical harmonic 
$U^{xyz}({\rm {\bf a}},{\rm {\bf b}},{\rm {\bf c}})$ relevant
for the angular dependence of dipolar transitions.} 
\begin{center}
\begin{tabular}{c}
$U^{110}={\rm {\bf a}} \cdot {{\rm {\bf b}}}$  \\
$U^{111}=\frac{2}{3}({\rm {\bf a}} \times {{\rm {\bf b}}})
\cdot {\rm {\bf c}}$  \\
$U^{112}=\frac{3}{2}({\rm {\bf a}} \cdot {{\rm {\bf c}}})
({\rm {\bf b}} \cdot {{\rm {\bf c}}})-\frac{1}{2}
({\rm {\bf a}} \cdot {{\rm {\bf b}}})$  \\
\end{tabular}
\end{center} 
\label{bipolar}
\end{table}
\noindent
dominated by the dipolar $3d\rightarrow 4f$
transitions. The electric dipolar and quadrupolar transition operators are
given by the Wigner-Eckart theorem:
\begin{eqnarray} 
\langle n| V_q |g\rangle &=&
P_{cl} \sum_{\gamma\lambda\sigma} 
(-)^{l-\lambda}\left (
\begin{array}{ccc}
l & Q & c \\
-\lambda & q & \gamma \\
\end{array}
\right )
\langle n| l_{\lambda\sigma}^{\dagger}c_{\gamma\sigma} |g \rangle ,
\nonumber
\end{eqnarray} 
where $l_{\lambda\sigma}^{\dagger}$ creates an electron in shell
$l$ with orbital component $\lambda$ and spin $\sigma$. 
The reduced matrix element is given by
\begin{eqnarray} 
P_{cl}(Q)=(-1)^l n_{lQc} [lc]^{\frac{1}{2}} 
\int dr R_{n_l l}(r)r^Q R_{n_c c}(r) .
\end{eqnarray} 

For rare earths the crystal fields on the $4f$ electrons
 are often negligibly small and one can assume spherical
symmetry ($SO_3$). For a magnetic system the symmetry is 
lowered to $SO_2$ and the $J$-values branch into $M_J=-J,\dots,J$.
Here we take the magnetic axis along the ${\hat {\rm {\bf Z}}}$-axis and the 
ground state to be $M_J=J$. To obtain a non-zero intensity
the total transition operator has to be totally symmetric.
This means we have to consider $I^z_0(\omega)$ for XAS and
$I^{zz'r}_0(\omega,\omega')$ for RIXS.
$I^z_0$ is a combination of the matrix elements
$I_q(gnng)=\langle g |V_q^{\dagger} |n\rangle \langle n|V_q |g \rangle$.
The combinations for $Q=1,2$ are given in Table \ref{fund}. For
dipolar transitions we have the well-known spectra: isotropic
($I^0_0$), circular dichroic ($I^1_0$), and linear
dichroic ($I^2_0$).

\begin{figure}[h]
\centering
\epsfxsize 8.0cm
\leavevmode
\epsfbox{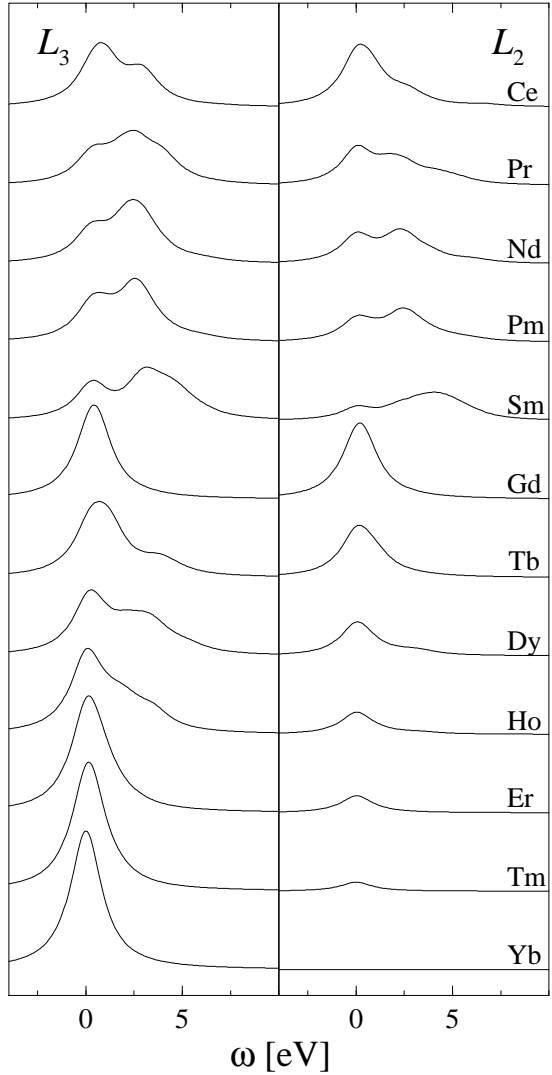}
\caption{Isotropic quadrupolar $2p\rightarrow 4f$ XAS spectra 
as a function of excitation energy $\omega$ for trivalent rare-earth ions. 
The left and right panel show the $L_3$ and $L_2$ edge, respectively. 
Spectra for the $L_2$ edge have been scaled by a factor two.}
\label{Lz0}
\end{figure}

\begin{figure}[h]
\centering
\epsfxsize 8.0cm
\leavevmode
\epsfbox{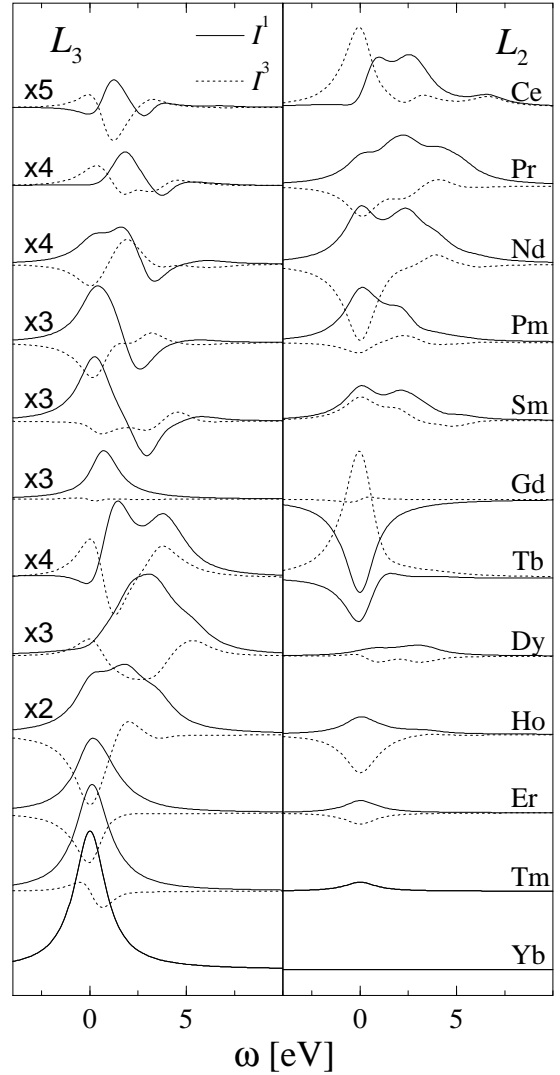}
\caption{
Quadrupolar $2p\rightarrow 4f$ XAS spectra 
as a function of excitation energy $\omega$ for trivalent rare-earth ions. 
The spectrum for left minus right circularly polarized light is 
a angular dependent combination of $I^1$ (solid line) and $I^3$ (dotted line),
see text.
The left and right panel show the $L_3$ and $L_2$ edge, respectively. 
Spectra for the $L_2$ edge have been scaled by a factor two.
Furthermore, some of the spectra are scaled relative to the spectra 
of  Fig. {\protect {\ref{Lz0}}}. The scaling also applies to the 
right panel.}
\label{Lz13}
\end{figure}

\begin{figure}[h]
\centering
\epsfxsize 8.0cm
\leavevmode
\epsfbox{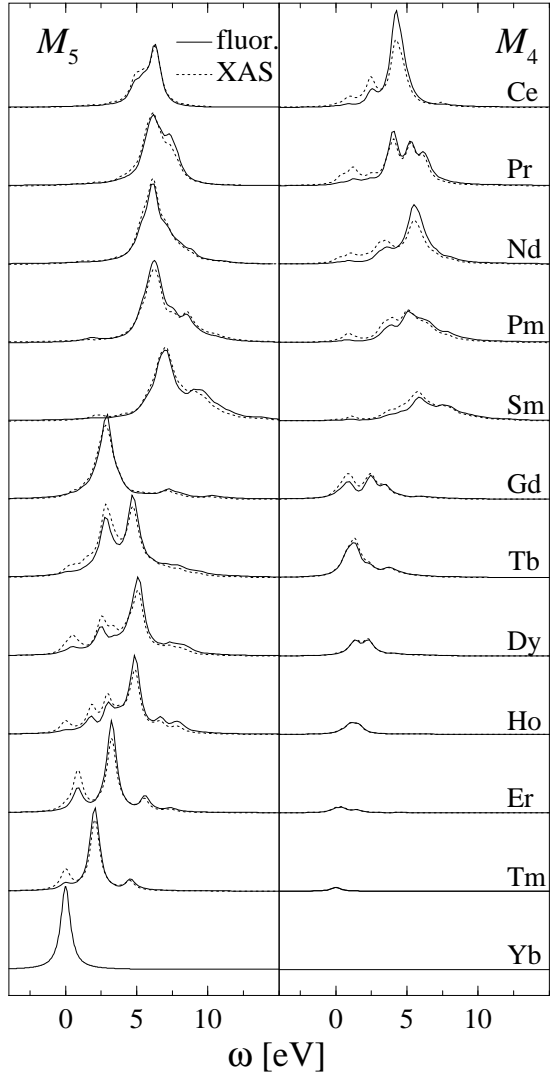}
\caption{Isotropic dipolar XAS spectra  
($4f^n\rightarrow \underline{3d}4f^{n+1}$, dotted line)
and fluorescence spectra for isotropic incoming and outgoing light
($4f^n\rightarrow \underline{3d}4f^{n+1}\rightarrow 4f^n$, solid line)
as a function of excitation energy $\omega$ for trivalent rare-earth ions. 
XAS and fluorescence have been normalized to  the intensity 
integrated over both edges; the scaling between spectra for different ions
is arbitrary.
The left and right panel show the $M_5$ and $M_4$ edge, respectively. }
\label{Mz0}
\end{figure}

\begin{figure}[h]
\centering
\epsfxsize 8.0cm
\leavevmode
\epsfbox{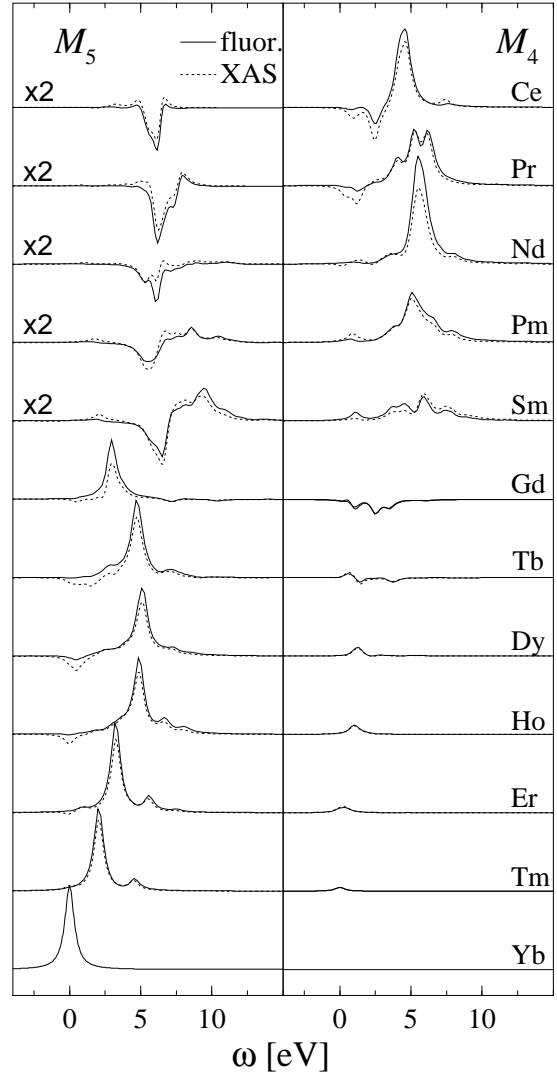}
\caption{
Same as Fig. \protect{\ref{Mz0}} but now for left minus
right circularly polarized incoming light.
Some of the spectra are scaled relative to the spectra 
of  Fig. \protect{\ref{Mz0}}. The scaling applies also to the 
right panel.}
\label{Mz1}
\end{figure}

Figures \ref{Lz0}-\ref{Mz1} give the XAS spectra for transitions
into the $4f$-shell at the $L_{2,3}$ and $M_{4,5}$ edges.
Calculations were done in the atomic limit using Cowan's
programs.\cite{Cow} The Hamiltonian includes 
the Coulomb interactions in the $4f$ shell and those 
between the $4f$ shell and the core hole and the spin-orbit coupling.
Parameters were obtained in the Hartree-Fock limit and 
the values for the Coulomb interaction were scaled down to 80 \% to account
for screening effects. The zero of the energy scale
corresponds to the energy of the lowest eigenstate in a spin-orbit
manifold.
Figure \ref{Lz0} gives the isotropic $2p \rightarrow 4f$ XAS
spectrum. The used Lorentzian of 2 eV is smaller than the 
expected $2p$ lifetime broadening. This was done since recent experiments
by Loeffen {\it et al.}\cite{Loef} 
demonstrate the  possibility of partial deconvolution 
of the lifetime broadening. By deconvoluting high-quality XAS data
by a Lorentzian with a width of 3 eV, they obtained a clear separation
of dipolar and quadrupolar features. Also the spectral line shape
of the pre-edge structures was in agreement with multiplet
calculations. This method seems to be well suited to 
\begin{table}
\caption{Coeficients ${\cal T}_z$ defined in Eqn.(\protect{\ref{Tzeqn}})
for $Q$=1,2 and $z=0,\dots,2Q$.  } 
\begin{center}
\begin{tabular}{ccccccccc}
&  $z=$ & 0 & 1 & 2 &  3 & 4 \\ [1.0mm] 
\hline \\
$Q=1$ &    & 1 & $-\frac{3}{2}$ & $\frac{1}{2}$ &   &  \\
~ ~ ~ 2 &  & 1 & $-1$ & $-\frac{5}{7}$ & 1 &  $-\frac{2}{7}$
\end{tabular}
\end{center} 
\label{Tz}
\end{table}
\noindent
study
the relative contributions of quadrupolar and dipolar transitions.
The spectrum obtained after taking the 
difference between left
and right circularly polarized light
\begin{table}
\caption{The spectra $I^{z}_0$ expressed in $I_q$ for $Q$=1,2.} 
\begin{center}
\begin{tabular}{ccc}
&$ I^{0}_0 = I_1 + I_0 + I_{-1}$&\\
$Q$=1 (dipolar)&$ I^{1}_0 = I_1 - I_{-1}$&\\
&$ I^{2}_0 = I_1 -2 I_0 + I_{-1}$&\\ \\
&$ I^{0}_0 = I_2 + I_1 + I_0 + I_{-1} + I_{-2}$& \\
&$ I^{1}_0 = I_2 +\frac{1}{2} I_1 -\frac{1}{2} I_{-1} - I_{-2}$& \\
$Q$=2 (quadrupolar)&
$ I^{2}_0 = I_2 -\frac{1}{2} I_1 - I_0 - \frac{1}{2}I_{-1} + I_{-2}$& \\
&$ I^{3}_0 = I_2 -2 I_1  + 2 I_{-1} - I_{-2}$ &\\
&$ I^{4}_0 = I_2 -4 I_1 + 6 I_0 - 4 I_{-1} + I_{-2}$ &\\
\end{tabular}
\end{center} 
\label{fund}
\end{table}
\noindent
is a combination of 
$I^z_0$ with odd $z$. Using Eqn. (\ref{Tzeqn}) we obtain
\cite{PCAl}
\begin{eqnarray} 
\frac{1}{2} \{
I({\hat {\rm{\bf k}}}\makebox{\boldmath {$ \epsilon$}}^+ \omega) -
I({\hat {\rm{\bf k}}}\makebox{\boldmath {$ \epsilon$}}^- \omega)\}
&=& \frac{2\pi}{\hbar}
N_{\omega}B_2^2 \{ -C^1_0({\hat {\rm {\bf k}}}) I^1_0(\omega) 
\nonumber \\ && \qquad  +
C^3_0({\hat {\rm {\bf k}}})  I^3_0(\omega) \} ,
\end{eqnarray}
where $C^z_0({\hat {\rm {\bf k}}})$ are Legendre polynomials of 
order $z$, i.e., $P^z(\cos \theta)$ with $\theta$ the angle between 
the direction of the 
light and the magnetic axis. The spectra $I^1_0(\omega)$
and $I^3_0(\omega)$ are given in Fig. \ref{Lz13}. 
\noindent

Figures \ref{Mz0} and \ref{Mz1} show, respectively, the isotropic
and circular dichroic $M_{4,5}$ XAS spectra
($\Gamma$=0.8 eV).
These spectra have been published earlier\cite{Thre,Goed,ImJo}
and are included for comparison with the fluorescence yield spectra.

The behaviour of the intensities can be related to ground state
expectation values by sum rules.\cite{Thsum,PCsum,ThBR}
For $l=c+Q$ one can reduce the expressions to\cite{Thmom,MvVRRS}
\begin{eqnarray} 
I^z_0(j)= \int_j d \omega I^z_0(\omega)
=\frac{P_{cl}^2}{[cl]}\sum_{xy} M_y(j) N_{xyz} 
\langle w^{xyz}_0 \rangle
\label{sumedge}
\end{eqnarray} 
with $M_y(j)\mathopen{=}c+1,c,c,-c$ for 
$jy\mathopen{=}j^+ 0,j^+ 1,j^- 0,j^- 1$ 
\noindent
($j^{\pm}=l\pm s$) and $N_{xyz}\mathopen{=}1,\frac{z}{[z]},\frac{z+1}{[z]}$ for
$xyz=z0z; z-1,1,z; z+1,1,z$.
The coupled tensor operators are defined as\cite{Thmom,Judd}
\begin{eqnarray} 
w^{xyz}_{\zeta} &=& \sum_{\lambda\lambda'\sigma\sigma'\xi\eta}
(-)^{l-\lambda'+s-\sigma}
\left (
\begin{array}{ccc}
l & x & l \\
-\lambda' & \xi & \lambda \\
\end{array}
\right )
\left (
\begin{array}{ccc}
s & y & s \\
-\sigma' & \eta & \sigma \\
\end{array}
\right )
\nonumber \\ &\times&
(-)^{z-\zeta}
\left (
\begin{array}{ccc}
x & y & z \\
\mathopen{-}\xi & \mathopen{-}\eta & \zeta \\
\end{array}
\right )
l_{\lambda'\sigma'}l_{\lambda\sigma}^{\dagger}~ 
n_{lx}^{-1} n_{sy}^{-1} {\underline n}_{xyz}^{-1}
\end{eqnarray} 
with $s\mathopen{=}\frac{1}{2}$. The operators are
spin independent and dependent for  $y=0$ and 1, respectively. 
These hole coupled 
tensor operators are related to the 
electron operators defined by Carra {\it et al.}\cite{PCPhB} via
$O^{xyz}\mathopen{=}2[l]\delta_{x,0}\delta_{y,0}
-(-1)^z r_{lx} r_{sy}w^{xyz}_0$
with $r_{lx}=\frac{(2l)!}{2^x (2l\mathopen{-}x)!}$. 
The advantage of using normalized operators
over Judd's operators\cite{Judd} is 
\begin{figure}[h]
\centering
\epsfxsize 7.5cm
\leavevmode
\epsfbox{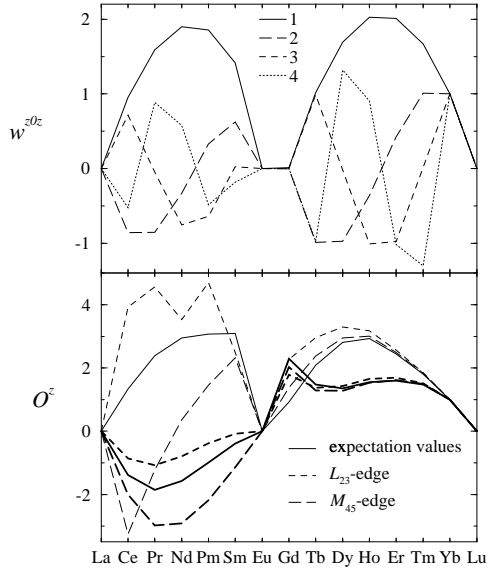}
\caption{The upper panel shows the ground state expectation
values of $w^{z0z}$ for trivalent rare-earth ions for $z$=
1 (solid line), 2 (long-dashed), 3 (dashed), and 4 (dotted).
The lower panel shows the ground state expectation value of
$O^z=\frac{z}{[z]}w^{z-1,1,0}+\frac{z+1}{[z]}w^{z+1,1,z}$ (solid line).
The dashed and long-dashed lines show 
$n_h[I^z_0(j^+)-\frac{c+1}{c}I^z_0 (j^-)]$
for the $L_{23}$ and $M_{45}$ edges, respectively. This quantity is
equal to $O^z$ according to the sum rules, see text.
The thin and thick lines correspond to $z$=0 and 1, respectively.
}
\label{oper}
\end{figure}\noindent
that the square roots
are removed from the expressions. They are normalized in such a 
way that the expectation values are unity for a ground 
state with one hole and $M_J=J=l+\frac{1}{2}$ (i.e., for
rare earths the ground state of Yb$^{3+}$, 
$f^{13}$($^2 F_{\frac{7}{2}}; M_J=\frac{7}{2})$).
Some typical examples are
the number of holes: $n_h=w^{000}_0$, the orbital 
and spin magnetic moments: $L_z =lw^{101}_0$ and
$S_z =sw^{011}_0$, respectively, and the spin-orbit coupling:
${\rm {\bf L}}\cdot{\rm {\bf S}} =-lsw^{110}_0$.
Note that the relation for $O^{xyz}$ does not hold for 
the magnetic dipole operator $T_z =\frac{l}{2l+3}w^{211}_0$
due to its definition in real instead of angular momentum space.

By summing Eqn. (\ref{sumedge}) over both spin-orbit split  edges,
\begin{eqnarray} 
{\overline I}^z_0 (j^+) +
{\overline I}^z_0 (j^-)
=\frac{\langle w^{z0z}_0 \rangle}{\langle n_h \rangle}
\label{sumadd}
\end{eqnarray} 
we obtain the sum rule 
that relates the total integrated intensity to
the spin-independent operators.\cite{Thsum}
 (${\overline I}$ indicates
that the spectra are normalized to the isotropic spectrum).
For $z=1$ we have the ``$L_z$'' sum rule.
In general, it relates the integrated intensity of $I^z_0(\omega)$ to 
spin-independent operators that give the $2^z$-polar moment 
in the electron distribution. For $z=0$ it simply says that the integrated
intensity over both edges is proportional to the number of holes.
For higher values of $z$ the ground state expectation
values of $w^{z0z}$ show an oscillatory 
behaviour along the rare earth series, see Fig. \ref{oper}.
This behaviour is already well described by assuming a 
Hund's rule ground state.\cite{MvVRRS}

A weighted substraction of both edges can be related to the spin-dependent
operators,
\begin{eqnarray} 
{\overline I}^z_0 (j^+) &-& \frac{c+1}{c}
{\overline I}^z_0 (j^-)
=\nonumber \\ &&
\frac{1}{\langle n_h \rangle}
\left \{
\frac{z}{[z]} \langle w^{z\mathopen{-}1,1,z}_0 \rangle
+\frac{z+1}{[z]} \langle w^{z\mathopen{+}1,1,z}_0 \rangle
\right \}
\end{eqnarray} 
For $z=0$ this expression forms the basis of the theory 
of branching ratios.\cite{ThBR} It relates the integrated intensities
of the spin-orbit split edges of the isotropic spectrum
to the ground state expectation value of the spin-orbit 
operator (note that $w^{-1,1,0}=0$). For $z=1$ it gives the 
``$S_z$-$T_z$'' sum rule.\cite{PCsum} 

The assumption in deriving this 
equation is that a certain spin-orbit split edge can be described by
the core hole having the corresponding $j$-value.
This approximation is valid for late rare earths (and also 
late transition metals). For early rare earths 
deviations are found for the spin-dependent sum rule, see Fig. \ref{oper}. 
Although for these systems a smaller spin-orbit
coupling is found, this is not the dominant effect that 
causes the stronger mixing of the two edges. More important 
are the $LSJ$ values that are reached by the absorption. For late rare earths 
the ground state has maximum $LSJ$ values.
 As a result of the dipole 
selection rules ($\Delta S=0,\Delta L=0,\pm 1$) also high $LSJ$
values for XAS final states are found. These states are 
predominantly found in the $j^+$-edge. A qualitative argument
for this goes as follows. Within $LS$-coupling (i.e.,
no spin-orbit coupling) the ${\underline c}l^{n+1}$ states
with high $LS$ values have relatively low energies (where
${\underline c}$ indicates a core hole). After switching
on the spin-orbit coupling the $LS$ states with low energy 
predominantly go into the $j^+$-edge and those with high energy
into the $j^-$-edge. Hence, for the states with high $LSJ$ values
there is only small mixing between the two edges. The most
extreme examples are the states with maximum $LSJ$ values that
only occur in the $j^+$-edge. The limited presence
of high $LSJ$ character in the $j^-$ edge directly explains why the 
the $L_2$ and $M_4$ edge have little intensity for late 
rare earths.

The situation is different for early rare earths. First, the
total $J$ of the ground state is given 
by $|L-S|$. Second, for early rare earths
the maximum spin of the XAS final states is higher than that
of the ground state ($S_n^{\rm max}=S_g^{\rm max}+1$).\cite{ThBR}
Since in XAS $\Delta S=0$ these states have little weight
(as a result of the mixing by the spin-orbit coupling 
they obtain a finite intensity). For early rare earths excitations
therefore occur at intermediate $LSJ$ values. This can
also be seen in the $M_{45}$ XAS spectra where for early rare earths
significant intensity only occurs
at 5-10 eV above the absorption edge (which consists 
of the ``dipole-forbidden'' maximum spin states), see Fig.
\ref{Mz0}. The stronger mixing between the edges is directly
apparent from strong intensity at both edges. This mixing
has as a result that for early rare earths deviations 
for the spin-dependent sum rule are found, 
although for $z=1$ the trends are well predicted.

\subsection{$2p\rightarrow 5d$ XAS and elastic scattering}  
We now discuss the behaviour of XAS and resonant elastic scattering
at the $2p\rightarrow 5d$ absorption edge. For elastic scattering,
it is more convenient to couple the matrix elements of the 
excitation to those of the deexcitation instead of coupling the 
excitation and deexcitation to their complex conjugates, as was done 
for inelastic scattering, see Eqn. (\ref{RIXS}).
One then obtains for dipolar transitions
\begin{eqnarray} 
I({\rm{\bf k}}\makebox{\boldmath {$ \epsilon$}} \omega,
{\rm{\bf k}}'\makebox{\boldmath {$ \epsilon$}}' \omega)
\mathopen{=}\frac{2\pi}{\hbar} N_{\omega}^2 
 | \sum_{z,n} \frac{B_1^2}{{\cal E}_n}
 {\rm {\bf T}}^{z} (\makebox{\boldmath {$ \epsilon$}},
\makebox{\boldmath {$ \epsilon$}}')
\cdot 
{\rm {\bf I}}^{z}(gnng)
|^2
\nonumber
\end{eqnarray} 
where the angular dependence ${\rm {\bf T}}^{z} (
\makebox{\boldmath {$ \epsilon$}},\makebox{\boldmath {$ \epsilon$}}')$
is given by Eqn. (\ref{Tzdip}) under the replacement
of $\makebox{\boldmath {$ \epsilon$}}^*$ by
$\makebox{\boldmath {$ \epsilon$}}'^*$. 
For XAS, the $z=1$ spectrum is selected by taking the difference between
left and right circularly polarized light. In elastic scattering
one adopts a $\sigma\rightarrow \pi$ scattering geometry, where
$\sigma$ and $\pi$ denote linear polarization perpendicular to and in the
scattering plane, respectively. 
From the expressions for $U^{11z}(\makebox{\boldmath {$ \epsilon$}}_{\sigma},
\makebox{\boldmath {$ \epsilon$}}_{\pi},{\hat {\rm {\bf Z}}})$,
see Table \ref{bipolar},
one sees that the $z=1$ contribution is obtained if 
${\hat {\rm {\bf Z}}}$, i.e., the magnetic axis, is in the scattering plane.
Thus the $\sigma\rightarrow\pi$
spectrum can be approximately considered as the square of the
circular dichroic XAS spectrum.

The difficulty in the interpretation of the $2p\rightarrow 5d$
circular dichroism is that the radial matrix elements are strongly affected
by the presence of $4f$ electrons.\cite{Harm} 
Since the radial extent of the $5d$ orbital depends on the direction
of the $5d$ moment with respect to that of the $4f$ (and therefore
on energy) deviations from the XAS sum rules should be found. 
This was indeed observed experimentally where the sign of the 
signal could not be reconciled with the expected direction of the
magnetic 
moment.\cite{Baud,Gior,Gior2} The change in radial matrix element was 
first considered by band structure calculations.\cite{Harm}
However, in LDA the circular dichroic 
branching ratio is always $-1$, since
only the $5d$-$4f$ spin interactions are considered.
Recently, a model has been developed that includes the complete
$df$-Coulomb interaction, i.e., also the orbital contributions.\cite{MvVl23}
The assumption was that the dichroic signal was mainly determined
by the change in the radial matrix elements and that the $5d$ polarization
in the ground has a relatively small effect, i.e., one can assume
a $5d^0$ ground state. For constant radial matrix elements,
the integrated intensity of the circular 
dichroism is then zero.\cite{Thsum,PCsum}
However, this does not mean a zero signal. As a result of the interaction
with the $4f$ electrons the circular dichroic spectrum contains equal positive
and negative parts. Note, that this implies
a finite elastic scattering amplitude
even in the absence of a $5d$ polarization in the ground state.
Due to the large $2p$ lifetime broadening 
the spectrum for $q$-polarized light is mainly determined by its first moment,
$\langle g| V_q^{\dagger} H_{df} V_q |g\rangle$. 

Within a framework with fixed atomic orbitals one can create
a $5d$ orbital with a different radial extent by mixing 
\begin{figure}[h]
\centering
\epsfxsize 8.0cm
\leavevmode
\epsfbox{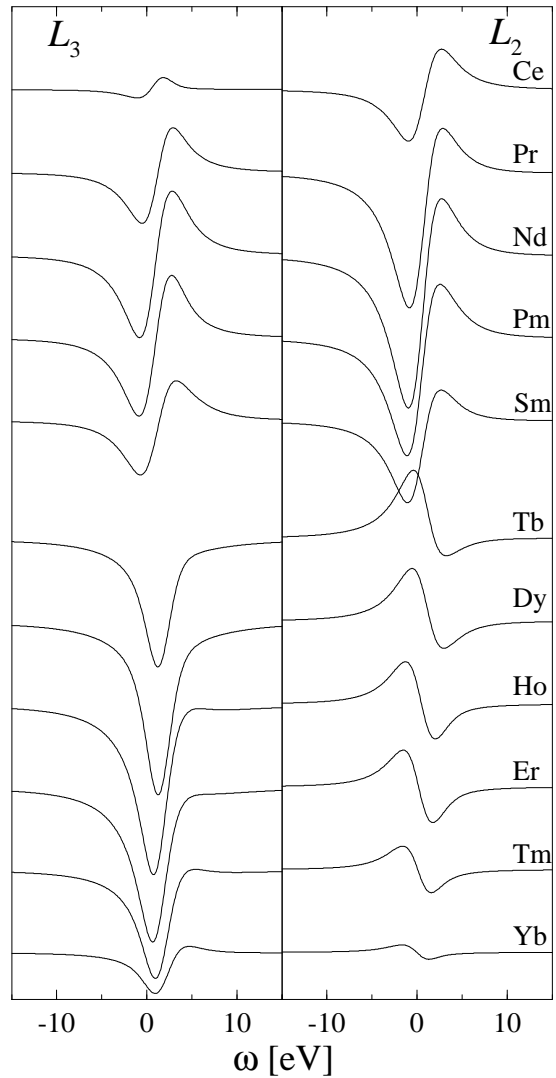}
\caption{Dipolar circular dichroic $2p\rightarrow 5d$
XAS spectra as a function of excitation energy $\omega$ for
trivalent rare-earth ions. The left and 
right panel show the $L_3$ and $L_2$ edge, respectively.
All spectra have the same scale.  }
\label{Lz15d}
\end{figure}\noindent
it with
other $nd$ orbitals, i.e., $|5{\tilde d}\rangle =a |5d\rangle
+\sum_{n\neq 5} a_n |nd\rangle$. Using first order perturbation 
theory one finds\cite{MvVl23} that the zeroth moment ${\tilde I}^z_0$ of the 
absorption spectrum into $5{\tilde d}$ is proportional to the first moment
$I^{(1)z}_0$ of the absorption spectrum into $5d$:
\begin{eqnarray}
{\tilde I}^z_0&=& -\sum_{n \neq 5} \frac{2}{\Delta_{nd}} I^{(1)z}_0(n,j)
\nonumber \\ &=&
- \sum_{n \neq 5} \frac{2 P_{5d} P_{nd}}{[cl] \Delta_{nd}} 
 \sum_k G^k_{5d,4f;4f,nd} 
\nonumber \\ &&\qquad \qquad\times
\sum_{xy} L_{kx}(df) M_y (j)N_{xyz} 
\langle w^{xyz}_0 \rangle  ,
\label{int5d}
\end{eqnarray}
where $\Delta_{nd}$ is the energy difference between the
$5d$ and a $nd$ orbital, and
\begin{eqnarray}
L_{kx}(df)=(-)^{x+k} n_{dkf}^2 \frac{[f]}{[c]} \frac{n_{fx}}{n_{dx}}
\left\{
\begin{array}{ccc}
f & f & k \\ d & d & x
\end{array}
\right\} .
\end{eqnarray}
The $F^k$ direct Coulomb terms have been omitted since they do not
contribute to the circular dichroic first moment. Note 
that the ground-state expectation values in Eqn. (\ref{int5d}) 
are those of the $4f$ and not the $5d$ shell. The result looks very
similar to the XAS sum rules of Eqn. (\ref{sumedge}). In particular,
for the $k$=1 terms one has $L_{kx}(df)=1/35$ and, 
except for a scaling, the same coefficients as for XAS
are found. 

Figure \ref{Lz15d} shows numerical calculations of the 
$2p\rightarrow 5d$ circular dichroism at the $L_{23}$ edges using 
final state configuration interaction
between $\underline{2p}4f^{n+1}5d^1$ and
$\underline{2p}4f^{n+1} nd^1$.
To reduce the size of the calculations only the $6d$ orbital
is included. To account for the other $nd$ orbitals
the Coulomb interactions between the $5d$ and the $6d$
shell are multiplied by four. The energy difference 
between the $5d$ and $6d$ orbitals
has been taken 15 eV; only the $5d$ region is shown in the figure.
The broadening is 6 eV which accounts for band effects and 
lifetime broadening. 
As a result of computer limitations we were unable
to obtain results for Gd$^{3+}$. One would expect the spectrum to
be given by a single negative peak at the $L_3$ edge and a
positive peak with equal intensity at the $L_2$ edge. 

Despite the crudeness of the model one finds qualitative 
agreement between theory and experiment.\cite{Baud,Gior,Gior2} 
For late rare earths the signal at the $L_3$ edge 
is larger than that at the $L_2$ edge, whereas for early rare earths
the $L_2$ is the larger. The sign of the $I^1_0$ quadrupolar contribution 
to the absorption spectrum is opposite to that of the dipolar, except
for the $L_2$ edge for $4f^9$ to $4f^{13}$ where the same
sign is found. This difference is an effect of the $G^3$ and $G^5$ terms
in Eqn (\ref{int5d}). Also the relative signs of the edges are 
well predicted. For early rare earths both edges have the same sign,
whereas for late rare earths an opposite sign is found. 

One also observes that, for early rare earths and for the $L_2$ edges
in late rare earths, the spectrum is more derivative-like than
for the $L_3$ edge in late rare earths. This effect is also seen 
experimentally.\cite{Baud,Gior} We do not have a satisfactory
qualitative explanation for this effect. We also note that for the $L_2$ edge
the signal occurs at lower energies with respect to the lowest final state.
This effect, combined with the larger lifetime broadening of the $L_2$-edge,
might explain why the quadrupole contributions are not always easily 
observed in the $L_2$ edge.

For finite $5d$ polarizations one can approximate the total intensity
by $I^z_0+{\tilde I}^z_0$, where $I^z_0$ is given by Eqn. (\ref{sumedge}).
Note that the $I^z_0$ and ${\tilde I}^z_0$ are given by $5d$ 
and $4f$ ground-state expectation values, respectively. 
Experiment shows that ${\tilde I}^z_0$ dominates. 
The trends in $|I^z_0|$ are similar to those of $|{\tilde I}^z_0|$,
but their  signs are usually opposite.\cite{JoIm}
A finite $5d$ polarization therefore
decreases the integrated
intensity (in principle it could even change sign),\cite{Mats}
but has relatively little effect on  the branching ratio.

\subsection{Resonant inelastic x-ray scattering}
For x-ray inelastic scattering sum rules one has to integrate over
the incoming and outgoing photon energies. 
Let us first consider the integration
over the transferred energy, $\Delta \omega=\omega-\omega'$, which leads to a 
term describing the decay of an intermediate state,
$I^{z'}_{\zeta'}(n'n)=\sum_f I^{z'}_{\zeta'} (n'ffn)$.
Note the presence of cross terms as a result of interference 
between intermediate states leading to the same final state.
For the decay one has to distinguish two situations.

First, the core hole is filled by an electron from a shallower
core level, leading to a Raman process described by
$l^n \rightarrow {\underline c} l^{n+1}
\rightarrow {\underline c}' l^{n+1}$. If both the intermediate
and final states are split by the spin-orbit coupling 
the spectrum has four clearly separated manifolds.
For a given manifold we find\cite{MvVRRS}
\begin{eqnarray} 
I^{z'}_{\zeta'}&& (nn';jj')= 
{\cal B}^{z'}_{jj'}(c,Q,c')
\sum_{mm'}
(-)^{j-m'}
\left (
\begin{array}{ccc}
j & z' & j \\
-m' & \zeta' & m \\
\end{array}
\right )
\nonumber \\  &&\qquad\qquad\qquad\qquad\qquad \times
\langle n' |c_{jm'}c_{jm}^{\dagger} | n\rangle  
n_{jz'}^{-1} ~,
\label{corepol}
\end{eqnarray} 
with the coefficient given by
\begin{eqnarray} 
{\cal B}^{z'}_{jj'}(c,Q,c') &=&
(-)^{j+j'+c+c'} P_{cc'}^2 [jj']
\left \{
\begin{array}{ccc}
j & j' & Q' \\
c' & c &  s
\end{array}
\right \} ^2
\nonumber \\  &&\qquad\times
\left \{
\begin{array}{ccc}
Q' & Q' & j' \\
j & j &  z'
\end{array}
\right \} 
n_{jz'} n_{Q'z'}^{-1} ~ .
\end{eqnarray} 
The decay of an intermediate state ($n=n'$)
is therefore  determined by an expectation value
of the polarization of the core hole for that state. 
For, e.g., $z'=1$ this expectation value is $\langle j_z \rangle$.

\begin{figure}[h]
\centering
\epsfxsize 7cm
\leavevmode
\epsfbox{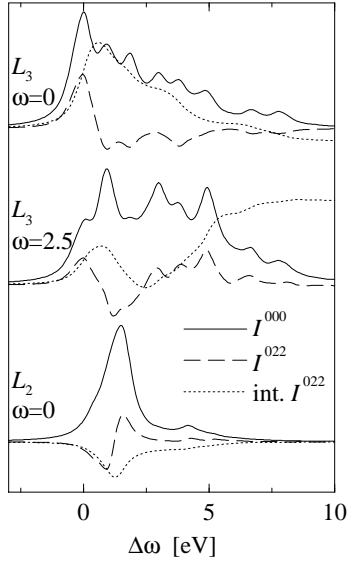}
\caption{RIXS spectra ($4f^n \rightarrow 
\underline{2p}4f^{n+1} \rightarrow \underline{3d}4f^{n+1}$)
as a function of transferred energy
$\Delta \omega=\omega-\omega'$. The zero of the energy scale corresponds to
the energy of the lowest final state of a certain edge.
Spectra are given at three 
excitation energies, from top to bottom: at the $L_3$ absorption
edge, 2.5 eV above the $L_3$ absorption edge, at the $L_2$ absorption
edge, see Fig. \protect{\ref{Lz0}}. Only the $M_5$ edge is shown. 
The incoming light is isotropic;
the outgoing light is isotropic 
($I^{000}_0(\omega,\omega')$, solid line) and
linear dichroic ($I^{022}_0(\omega,\omega')$,
dashed line). The dotted line gives the intensity of 
$I^{022}_0(\omega,\omega')$
integrated along the $\Delta \omega$ axis.}
\label{HoRRS}
\end{figure}

Let us consider a specific example. 
A typical resonant Raman process\cite{Ham,Kri} is given by
$l\mathopen{=}4f$, $c\mathopen{=}2p$, and $c'\mathopen{=}3d$.
For simplicity we take the incoming light isotropic.
If the system is magnetic the absorption process 
still creates a polarized core hole as a result of the polarization of the 
valence shell. 
If the polarization is not detected the outgoing light has
an isotropic and linear dichroic part, see Eqn. (\ref{Tzeqn}),
\begin{eqnarray} 
\frac{1}{2} \{
&&I( \omega, {\hat {\rm{\bf k}}}'\makebox{\boldmath {$ \epsilon$}}^+ \omega') +
I(\omega, {\hat {\rm{\bf k}}}'\makebox{\boldmath {$ \epsilon$}}^- \omega')\}
\nonumber \\  && \quad
= \frac{2\pi}{\hbar}
N_{\omega'} N_{\omega}B_1^2 B_2^2 \{  I^{000}_0(\omega,\omega') +
\frac{1}{2} C^2_0({\hat {\rm {\bf k}}})  I^{022}_0(\omega,\omega') \} .
\nonumber
\end{eqnarray}
Figure \ref{HoRRS} gives the two different spectra for Ho$^{3+}$
at three different incoming photon energies.
If we now take, for a certain absorption energy,
the difference of two spectra at different detection
angles with respect to the magnetic axis,
we obtain a signal that is proportional to $I^{022}_0(\omega,\omega')$. In that case the integration over $\omega'$
of a spectrum at a given absorption energy is finite at the
$L_3$-edge, but zero at the $L_2$-edge, see Fig. \ref{HoRRS}.
This is a direct result
of the fact that the $j=\frac{1}{2}$ level has no quadrupolar moment
(a finite intensity can still occur as a result of the presence
\noindent
of $j=\frac{3}{2}$ character in the $L_2$ edge due to the mixing by
the core-valence Coulomb interactions).
 
For a measurement with isotropic outgoing light the expectation value
is proportional to $\delta_{n,n'}$, i.e., the decay is constant
and all interference terms cancel. When a summation over both
$j'$-edges is made one obtains\cite{PCRRS,MvVRRS}
\begin{eqnarray} 
I^{z'=0}_0 (nn') &=&\sum_{j'} I^0_0 (nn';j j')=
\frac{ P_{cc'}^2}{[c]} \delta_{n,n'} ~.
\end{eqnarray}
Therefore, for isotropic outgoing light an integration over the 
transferred energy $\omega-\omega'$ leads to a spectrum as a function
of excitation energy $\omega$ which is proportional to 
XAS.\cite{PCRRS,MvVRRS} 

The second case is where the deexcitation involves the valence shell,
i.e., $l^n \rightarrow {\underline c} l^{n+1}\rightarrow l^n$.
Figures \ref{Mz0} and \ref{Mz1} show the spectra integrated over the
energy of the outgoing photons (fluorescence yield) for isotropic and
circular dichroic incoming light, respectively; the outgoing light
has been taken isotropic. We observe significant differences 
with the XAS spectra. Therefore,
even for isotropic outgoing light, the decay cannot be decoupled
from the excitation. In contrast to Eqn. (\ref{corepol}), where the operators
working on the closed shell could be removed, one obtains here
that the decay is proportional to an intermediate-state
expectation value of a two-particle operator. For isotropic outgoing
light one has\cite{MvVfl}
\begin{eqnarray} 
I^{z'=0}_0 (nn') &=&
\frac{ P_{cl}^2}{[c]} \{ \delta_{n,n'} +
\frac{1}{[l]n_{l1c}^2} \sum_{\lambda\lambda'\gamma\gamma'\sigma\sigma'}
\delta_{\lambda'-\gamma',\lambda-\gamma}
\nonumber \\ &\times&
c^1(c\gamma',l\lambda') c^1(c\gamma,l\lambda) 
\langle n' |  l_{\lambda\sigma} c_{\gamma'\sigma'}
 l_{\lambda'\sigma'}^{\dagger} c_{\gamma\sigma}^{\dagger}
|n \rangle \} ,
\nonumber
\end{eqnarray} 
where $c^1(c\gamma,l\lambda)$ are the Slater-Condon parameters.\cite{CSh}
The matrix elements are, except for an offset and
the radial integrals, equivalent
to those of the $G_{cl}^1$ Coulomb exchange. Note that the 
``$G_{cl}^1$'' term also assumes negative values and that the lowest
possible value for $I^{z'=0}_0$ is zero. The $G_{cl}^1$
Coulomb term is also for a large part responsible for the
position of the eigenstates in a spin-orbit manifold.
This directly explains the trend in the fluorescence spectra,
that the states closer to the absorption threshold in a certain
edge have in general a smaller decay compared to those at the 
high-energy side of an edge.

Let us now consider the sum rules for the intensities integrated 
along the transferred and excitation energy. For deep-lying core levels
with a large lifetime broadening, such as the $2p$ shell, one can 
use the fast-collision approximation.\cite{Luo,PCRRS}
 This implies replacing the
intermediate state energy denominator by ${\overline {\cal E}}_n=\omega+E_g-
{\overline E}_n + i\frac{\Gamma}{2}$, where ${\overline E}_n$
is an average energy. For the situation where the emission involves 
two core levels one can then derive\cite{PCRRS,MvVRRS}
\begin{eqnarray}
I^{zz'r}_0(j,j')&=&\int_j d \omega \int_{j'} d \omega' ~
I^{zz'r}_0(\omega,\omega') 
\nonumber \\ &=& {\cal B}^{z'}_{jj'} (c,Q,c')
\sum_{xy} {\cal C}^{xyrzz'}_j (c,Q,l) 
\langle w^{xyr}_0 \rangle
\end{eqnarray}
where the coefficients ${\cal B}$ and ${\cal C}$ are given 
in Ref. \onlinecite{MvVRRS}. For isotropic outgoing light 
we have ${\cal C}^{xyzz0}_j=\frac{P_{cl}^2}{[cl]}
M_y(j)N_{xyz}$, which means that the
same coefficients are found as for XAS. In this limit the fast-collision 
approximation is not necessary since all interference effects cancel.

When the deexcitation involves the valence states the situation
is more complex. Sum rules would involve two-particle 
valence shell expectation values which are difficult to evaluate.
Recently, an investigation has been made into the applicability of
XAS sum rules for spectra obtained with fluorescence yield.\cite{MvVfl}
Although fluorescence yield is in principle not equal to 
XAS the conditions for integrated intensities
are less stringent. Here one does not require that every final state
has a constant decay but that the total decay of the excited intermediate
states does not have a strong polarization dependence.

The state after excitation with $q$-polarized light can be written as
$|v_q\rangle =\sum_n a_n(q) |n\rangle$; this state can be normalized
to unity by using the coefficients 
${\overline a}_n(q)=a_n(q)/\sqrt{\langle v_q|v_q \rangle}$. The integrated
intensity of the fluorescence spectrum excited with $q$-polarized
light can then be written as\cite{MvVfl}
\begin{eqnarray}
I_q^{\rm fluor} =
\langle v_q |v_q \rangle \langle V^{\Gamma} \rangle_q
=I_q^{\rm XAS} \langle V^{\Gamma} \rangle_q ,
\end{eqnarray}
i.e, the integrated intensity is given by the XAS intensity multiplied by
a term that describes the radiative decay. If the polarization
of the outgoing light is not measured this latter term is given by
\begin{eqnarray}
\langle V^{\Gamma}\rangle_q =\sum_{z' ~ {\rm even}}
{\cal T}_{z'} C^{z'}_0({\hat {\rm {\bf k}}}') \sum_{nn'} 
\frac{\pi}{\Gamma}
\frac{ {\overline a}_n {\overline a}_{n'} }
{ \left (\frac{ E_n -E_{n'} }{\Gamma}\right )^2 +1 }
I^{z'}_0(nn') ~.
\nonumber
\end{eqnarray}
This clearly shows that the total decay for 
$q$-polarized light $\langle V^{\Gamma}\rangle_q$
is a weighted average of the decays of the intermediate states $I^{z'}_0(nn')$.
Proportionality of the integrated intensity of the circular dichroic
fluorescence spectra, i.e., 
$(\sum_q q I^{\rm fluor}_q) /(\sum_q I^{\rm fluor}_q)$,
with $\langle w^{101} \rangle = \langle L_z \rangle/l$ is obtained
if $\langle V^{\Gamma}\rangle_q$ is not strongly dependent 
on the polarization. Numerical evaluation of $\langle V^{\Gamma}\rangle_q$
shows that this situation is found for early rare earths, but 
that there is a strong polarization dependence 
for late rare earths.\cite{MvVfl}
This can be understood as follows. As was shown above, the decay of the 
intermediate states is proportional to a 
``$G_{cl}^1$''-like term. Maximum variations in decay can be expected
for ``pure'' $LS$-like intermediate states. The effect of mixing
by the spin-orbit coupling is small for the high $LSJ$ states
that are reached in absorption in late rare earths, leading
to a strong polarization dependence.
The stronger mixing of the intermediate $LSJ$-states that are reached in
early rare earths decreases variations in 
$\langle V^{\Gamma} \rangle_q $. It is remarkable that the 
mechanism that causes deviations for the ``$S_z$'' XAS sum rule
improves the agreement between $L_z$ values obtained by fluorescence 
yield and the ground-state expectation values.

It must be noted that a more efficient way to remove variations
in $\langle V^{\Gamma} \rangle_q $ is to create a $|v_q\rangle$
which has not a strong $LS$-like character. This is found for 
many transition-metal systems where a strong crystal field
quenches the orbital moment.  

\section{Conclusion}
In conclusion, a comparison has been made between XAS and resonant
inelastic x-ray scattering. 
Several aspects deserve further experimental investigation. 
The isotropic  $L_{23}$ spectra should be reexamined since, as was shown 
by Loeffen {\it et al.},\cite{Loef} more detailed information can be obtained
by partial deconvolution of the lifetime broadening. This
should enable a direct comparison of the relative size of
the dipolar and quadrupolar contributions. Also a careful 
determination of the intensities
of the two spin-orbit split edges would be interesting. It is generally
assumed that for isotropic light $I^0_{L_3}/I^0_{L_2}$= 2:1. 
However, a strong polarization of the $5d$ electrons in the ground
state would give deviations from the statistical  branching ratio.
The $2p\rightarrow 5d$ circular dichroism certainly requires more 
experimental and theoretical research. Explanation of circular dichoism
by band structure calculations
have so far only considered the $df$ spin interaction. The difficulties
in explaining  the circular dichroic branching ratios
directly implies that the coupling of the $5d$ band to the local 
$4f$ states is not well understood. 
Band structure calculations including orbital
polarization are necessary in the interpretation of the circular 
dichroism. 

The experimental developments in x-ray inelastic scattering
are more recent. It is now rather well established that on-resonance 
the excitation and decay cannot be decoupled. For a delocalized system
this implies that one has to take into account the crystal momentum 
of the core hole (although core-valence interactions in the intermediate
state might change its value). For a localized system one has to 
consider the angular momentum. Within an independent electron  model
the angular momentum is conserved. However, interactions
with the valence shell change the values of $j$ and $m$ and 
in general for an intermediate-state eigenstate one has to consider
expectation values of, e.g.,  $j_z$ of the core hole. This quantity 
could be obtained by, e.g., measuring the difference between  left and right
circular polarization of the outgoing light. 
Unfortunately these experiments are rather
complex. A determination of the quadrupolar moment of the 
core hole polarization
seems more promising since it involves a measurement at two different 
detection angles. The presence of core-hole polarization has been 
demonstrated in resonant photoemission\cite{ThRPES} ($l^n\rightarrow 
{\underline c} l^{n+1} \rightarrow {\underline c}'^2 l^{n+1} E_k$, where
$E_k$ denotes a photoelectron) which is formally very similar to the
scattering process $l^n \rightarrow {\underline c}l^{n+1} \rightarrow
{\underline c}' l^{n+1}$. Recent resonant Raman  experiments
on Co show the presence of higher moments in the 
core hole distribution.\cite{Brai}

The x-ray scattering process $l^n \rightarrow {\underline c}l^{n+1}
\rightarrow l^n$ enables one to study valence band excitations. Although
for electric multipole transitions $\Delta S=0$, these excitations
also include spin flips, since the spin is not a good quantum number in
the intermediate state as a result of the large 
core-hole spin-orbit coupling. Recently, octet-sextet transitions
have been observed in Gd$^{3+}$ (ground state $^8 S_\frac{7}{2}$).\cite{Dall}

A relatively simple way to study resonant inelastic x-ray scattering 
is fluorescence yield, since it does not involve the detection of the 
energy of the outgoing photon. The spectra for fluorescence yield already
show significant deviations from  the XAS cross section.
\appendix
\section{}
The tensor products in this paper are done with $3j$-symbols which
are up to a factor equivalent to those with Clebsch-Gordan coefficients  
\begin{eqnarray}
[{\rm {\bf a}}^{l}, {\rm {\bf b}}^{l'} ]^x_{\xi} 
&=& \sum_{\lambda\lambda'} a^{l}_{\lambda} b^{l'}_{-\lambda'} 
(-1)^{l-\lambda-\lambda'}
\left (
\begin{array}{ccc}
l & x & l' \\
-\lambda & \xi & \lambda'
\end{array}
\right ) 
\nonumber \\ &=&
(-1)^{l'}[x]^{-\frac{1}{2}}
\sum_{\lambda\lambda'} a_{\lambda}^l b_{\lambda'}^{l'}
C_{l\lambda, l'\lambda'}^{x\xi} ~ .
\end{eqnarray}
When more than one superscript is present the last one gives the
rank of the tensor; a tensor without superscript has rank one.
Furthermore, we often make use of the numerical factors
\begin{eqnarray}
n_{lx}=
\left (
\begin{array}{ccc}
l & x & l \\
-l & 0 & l
\end{array}
\right )
= \frac{(2l)!}{\sqrt{(2l-x)!(2l+1+x)!}}
\end{eqnarray}
and
\begin{eqnarray}
n_{xyz}=
\left (
\begin{array}{ccc}
x & y & z \\
0 & 0 & 0
\end{array}
\right ) .
\end{eqnarray}
The latter factor is zero for odd $x+y+z$. In certain cases
it is convenient to have a ``generalized'' form for $n_{xyz}$:
\begin{eqnarray}
{\underline n}_{xyz}=
&i^g& \left (
\frac{(g-2x)!(g-2y)!(g-2z)!}{(g+1)!}
\right ) ^{\frac{1}{2}}
\nonumber \\ &\times&
\frac{g!!}{(g-2x)!!(g-2y)!!(g-2z)!!}
\end{eqnarray}
with $g=x+y+z$. Note that $n_{xyz}$ is equal to ${\underline n}_{xyz}$
for even $g$. The purpose of these factors is to remove the 
square roots from the expressions. These square roots are a result
of the normalization of the $3j$-symbols which is inconvenient
when dealing with physical quantities, such as, operators and 
combinations of spectra. 

Some useful tensor products are:
the inner product, 
$[{\rm {\bf a}}^l , {\rm {\bf b}}^l ]^0_0 n_{l0}^{-1} 
\mathopen{=}{\rm {\bf a}}^l \mathopen{\cdot} {\rm {\bf b}}^l$,
the outer product, $[{\rm {\bf a}}, {\rm {\bf b}} ]^1 \mathopen{=}
-\frac{i}{\sqrt{6}}{\rm {\bf a}} \mathopen{\times} {\rm {\bf b}}$,
and the  coupling of two spherical harmonics of the same vector
$[{\rm {\bf C}}^{l}({\hat {\rm {\bf k}}}), 
{\rm {\bf C}}^{l'}({\hat {\rm {\bf k}}}) ]^x \mathopen{=}
{\rm {\bf C}}^x({\hat {\rm {\bf k}}}) (-)^l n_{ll'x}$ .
In this paper we also apply the orthogonality relation
$({\rm {\bf a}}^l \mathopen{\cdot} {\rm {\bf b}}^l)
({\rm {\bf c}}^{l'} \mathopen{\cdot} {\rm {\bf d}}^{l'})\mathopen{=}
\sum_x [x] [{\rm {\bf a}}^l, {\rm {\bf c}}^{l'} ]^x
\mathopen{\cdot} [{\rm {\bf b}}^l, {\rm {\bf d}}^{l'} ]^x$ with 
$[x]\mathopen{=}2x\mathopen{+}1$.


\begin{thebibliography}{99}
\bibitem{GSGd} G. Sch\"utz, W. Wagner, W. Wilhelm, 
P. Kienle, R. Frahm, and G. Materlik, Phys. Rev. Lett. {\bf 58}, 737 (1987).
\bibitem{GvdL} G. van der Laan, B. T. Thole, G. A. Sawatzky,
J. B. Goedkoop, J. C. Fuggle, J.-M. Esteva, R. C. Karnatak,
J. P. Remeika, and H. A. Dabkowska, Phys. Rev. B {\bf 34}, 6529 (1986). 
\bibitem{Gibbs} D. Gibbs, D. R. Harshman, E. D. Isaacs, D. B.
McWhan, D. Mills, and C. Vettier, Phys. Rev. Let. {\bf 61}, 
1241 (1988).
\bibitem{Hannon} J. P. Hannon, G. T. Trammell, M. Blume, and
D. Gibbs, Phys. Rev. Lett. {\bf 61}, 1245 (1988).
\bibitem{PCAl} P. Carra and M. Altarelli, Phys. Rev. Lett. {\bf 64},
1286 (1990).
\bibitem{PCGd} P. Carra, B. N. Harmon, B. T. Thole, M. Altarelli, and
G. A. Sawatzky, Phys. Rev. Lett. {\bf 66}, 2495 (1991).
\bibitem{Harm} X. D. Wang, T. C. Leung, B. N. Harmon, and P. Carra, Phys. Rev.
B {\bf 47}, 9087 (1993); J. C. Lang, S. W. Kycia, X. D. Wang, B. N. Harmon,
A. I. Goldman, D. J. Branagan, R. W. McCallum, K. D. Finkelstein,
{\it ibid.} {\bf 46}, 5298 (1992). 
\bibitem{AngDep} P. Fischer, G. Sch\"utz, S. St\"ahler, and G. Wiesinger,
J. Appl. Phys {\bf 69}, 6144 (1991); K. Shimomi, H. Maruyama,
K. Kobayashy, A. Koizumi, H. Yamazaki, and T. Iwazumi, Jpn. J.
Appl. Phys. {\bf 32-2}, 314 (1992); J. C. Lang, G. Srajer, C. Detlefs,
A. I. Goldman, H. K\"onig, X. Wang, B. N. Harmon, and R. W. McCallum 
Phys. Rev. Lett. {\bf 74}, 4935 (1995).
\bibitem{Ham} K. H\"am\"al\"ainen, D. P. Siddons, J. B. Hastings, and L. E.
Berman, Phys. Rev. Lett. {\bf 67}, 2850 (1991).
\bibitem{Kri} M. H. Krisch, C. C. Kao, F. Sette, W. A. Caliebe, 
K. H\"am\"al\"ainen, and J. B. Hastings, Phys. Rev. Lett. {\bf 74},
4931 (1995).
\bibitem{Loef} P. W. Loeffen, R. F. Pettifer, S. M\"ullender,
M. van Veenendaal, J. R\"ohler, and D. S. Sivia, Phys. Rev. B
{\bf 54}, 14 877 (1996).
\bibitem{Har} B. N. Harmon and A. J. Freeman, Phys. Rev. B {\bf 10}, 1979
(1974). 
\bibitem{JoIm} T. Jo and S. Imada, J. Phys. Soc. Jpn. {\bf 62},
3721 (1993). 
\bibitem{MvVl23} M. van Veenendaal, J. B. Goedkoop, and
B. T. Thole, Phys. Rev. Lett. {\bf 78}, 1162 (1997).
\bibitem{Thsum} B. T. Thole, P. Carra, F. Sette, and G. van der Laan,
Phys. Rev. Lett. {\bf 68}, 1943 (1992).
\bibitem{PCsum} P. Carra, B. T. Thole, M. Altarelli, and X. Wang,
Phys. Rev. Lett. {\bf 70}, 694 (1993).
\bibitem{Luo} J. Luo, G. T. Trammell, and J. P. Hannon, Phys. Rev. Lett. 
{\bf 71}, 287 (1993).
\bibitem{PCRRS} P. Carra, M. Fabrizio, and B. T. Thole, Phys. Rev. Lett.
{\bf 74}, 3700 (1995).
\bibitem{MvVfl} M. van Veenendaal, J. B. Goedkoop, and
B. T. Thole, Phys. Rev. Lett. {\bf 77}, 1508 (1996).   
\bibitem{VMK} D. A. Varshalovich, A. N. Moskalev, and V. K. Khersonskii,
{\it Quantum Theory of Angular Momentum} (World Scientific, Singapore,
1988).
\bibitem{ThIII} B. T. Thole and G. van der Laan, Phys. Rev. B {\bf 49},
9613 (1994).
\bibitem{Cow} R. D. Cowan {\it The Theory of Atomic Structure and Spectra}
(University of California Press, Berkeley, 1981).
\bibitem{Thre} B. T. Thole, G. van der Laan, J. C. Fuggle, G. A. Sawatzky,
R. C. Karnatak, J.-M. Esteva, Phys. Rev. B {\bf 32}, 5107 (1985).
\bibitem{Goed} J. B. Goedkoop, B. T. Thole, G. van der Laan,
G. A. Sawatzky, F. M. F. de Groot, and J. C. Fuggle,
Phys. Rev. B {\bf 37}, 2086 (1988).  
\bibitem{ImJo} S. Imada and T. Jo, J. Phys. Soc. Jpn. {\bf 59},
3358 (1990).
\bibitem{ThBR} B. T. Thole and G. van der Laan, Phys. Rev. A {\bf 38},
1943 (1988); Phys. Rev B {\bf 38}, 3158 (1988).
\bibitem{Thmom} B. T. Thole, G. van der Laan, and M. Fabrizio, Phys. Rev. B  
{\bf 50}, 11 466 (1994).
\bibitem{MvVRRS} M. van Veenendaal, P. Carra, and B. T. Thole,
Phys. Rev. B {\bf 54}, 16 010 (1996); Note that here of the 
reduced matrix elements $P_{cl}$ only $[cl]^{\frac{1}{2}}$ is maintained.
\bibitem{Judd} B. R. Judd, {\it Second Quantisation in Atomic Spectroscopy} 
(Johns Hopkins University Press, Baltimore, 1967).
\bibitem{PCPhB} P. Carra, H. K\"onig, B. T. Thole, and M. Altarelli,
Physica B {\bf 192}, 182 (1993).
\bibitem{Baud} F. Baudelet, Ch. Giorgetti, S. Pizzini, Ch. Brouder,
E. Dartyge, A. Fontaine, J. P. Kappler, and G. Krill,
J. El. Spec. {\bf 62} (1993) 153. 
\bibitem{Gior} Ch. Giorgetti, Ph. D. Thesis,
University of Paris-Sud (unpublished).
\bibitem{Gior2} C. Giorgetti {\it et al.}, Phys. Rev. B {\bf 48} 
(1993) 12 732.
\bibitem{Mats} H. Matsuyama, I. Harada, and A. Kotani,
accepted for publication in the J. Phys. Soc. Jpn.
\bibitem{CSh} E. U. Condon and G. H. Shortley, {\it The Theory
of Atomic Spectra} (Cambridge University Press, Cambridge, 1951).
\bibitem{ThRPES} B. T. Thole, H. A. D\"urr, and G. van der Laan,
Phys. Rev. Lett. {\bf 74}, 2371 (1995).
\bibitem{Brai} L. Braicovich {\it et al.}, (unpublished).  
\bibitem{Dall} C. Dallera, L. Braicovich, C. Ghiringhelli,
M. van Veenendaal, J. B. Goedkoop, and N. B. Brookes, accepted for
publication in Phys. Rev. B.
\end{thebibliography}
\end{document}